\documentclass[aps,prd,preprint,superscriptaddress,amsfonts,amssymb,amsmath,nofootinbib]{revtex4-1}
\usepackage{graphicx,bm,color,ulem,latexsym}
\usepackage[colorlinks,linkcolor=magenta,anchorcolor=blue,citecolor=blue]{hyperref}

\begin{document}

\title{Static and spherically symmetric solutions in $f(Q)$ gravity}
\author{Wenyi Wang}
\email{wangwy@mails.ccnu.edu.cn}
\affiliation{Institute of Astrophysics, Central China Normal University, Wuhan 430079, China}
\author{Hua Chen}
\email{huachen@mails.ccnu.edu.cn}
\affiliation{Institute of Astrophysics, Central China Normal University, Wuhan 430079, China}
\author{Taishi Katsuragawa}
\email{taishi@mail.ccnu.edu.cn}
\affiliation{Institute of Astrophysics, Central China Normal University, Wuhan 430079, China}

\begin{abstract}
$f(Q)$ gravity is the extension of symmetric teleparallel general relativity (STGR),
in which both curvature and torsion vanish, and gravity is attributed to nonmetricity.
This work performs theoretical analyses of static and spherically symmetric solutions with an anisotropic fluid for general $f(Q)$ gravity.
We find that the off-diagonal component of the field equation due to a coincident gauge leads to stringent restrictions on the functional form of $f(Q)$ gravity. 
In addition, although the exact Schwarzschild solution only exists in STGR, we obtain Schwarzschild-like solutions in nontrivial $f(Q)$ gravity and study its asymptotic behavior and deviation from the exact one.
\end{abstract}

\maketitle

\section{introduction}\label{Sec: introduction}

Modern observations have confirmed that the expansion of the current Universe is accelerating,
which could be sourced by some unknown components called dark energy (DE)~\cite{Riess:1998cb,Planck:2018vyg}.
A possible way to explain DE is to modify pure general relativity (GR),
where gravity is ascribed to curvature, with the torsion and nonmetricity assumed to vanish.
For instance, the Lambda cold dark matter model is the minimal modification, 
where the cosmological constant $\Lambda$ is responsible for DE.
Along this way, one can generalize GR to $f(R)$ gravity by improving the Ricci scalar $R$ to its functional form~\cite{DeFelice:2010aj,Nojiri:2017ncd}.
On the other hand, choosing the torsion $T$ or the nonmetricity $Q$ as a geometric basis provides two different but equivalent descriptions of gravity.
These are the so-called teleparallel equivalent of general relativity (TEGR)~\cite{Hayashi:1979qx,Maluf:2013gaa}
and symmetric teleparallel general relativity (STGR)\footnote{
This theory is also called symmetric teleparallel equivalent to general relativity.
}
~\cite{Adak:2004uh,Adak:2005cd,Adak:2008gd,BeltranJimenez:2017tkd}.

In STGR theory, both the curvature and the torsion vanish, as the nonmetricity describes the gravity.
In this theory, under the teleparallelism constraint, we can always choose the coincident gauge, which restricts the affine connection to disappear and makes the metric tensor the only basic variable.
In analogy to $f(R)$ gravity, TEGR and STGR can be generalized to $f(T)$ gravity~\cite{Cai:2015emx,Bahamonde:2021gfp} and $f(Q)$ gravity~\cite{Heisenberg:2018vsk,BeltranJimenez:2017tkd}.
Although the latter one is relatively less investigated, it has many similar properties to those in $f(T)$ gravity.
For instance, similar to $f(T)$ theory~\cite{Li:2010cg,Sotiriou:2010mv,Tamanini:2012hg},
the gauge choice breaks the coordinate transformation invariant in $f(Q)$ theory,
which predicts different consequences in various coordinate systems~\cite{Zhao:2021zab}.
Recently, there have been several applications of $f(Q)$ theory:
cosmology~\cite{BeltranJimenez:2019tme,Anagnostopoulos:2021ydo,Barros:2020bgg,Dialektopoulos:2019mtr},
bouncing model~\cite{Bajardi:2020fxh}, 
wormhole solutions~\cite{Hassan:2021egb}, 
energy conditions~\cite{Mandal:2020lyq},
and the Newtonian limit~\cite{Flathmann:2020zyj} have been discussed.

In analogy to $f(T)$ gravity,
it is known that the nonzero off-diagonal component of field equations in $f(T)$ gravity,
which originates from the specific gauge choice,
restricts the functional form of $f(T)$~\cite{Ferraro:2011ks}.
Therefore, it would also put restrictions on the functional form of $f(Q)$ gravity,
which potentially gives us a guideline for the model building of $f(Q)$ gravity.
This work aims to investigate possible functional forms of $f(Q)$ under the restriction of the static and spherically symmetric geometry with an anisotropic fluid. 
In particular, we will show that there is no exact Schwarzschild solution for the nontrivial $f(Q)$ function. 
With the nonmetricity scalar $Q$ being constant, 
we also analyze the deviation of the metric from the exact Schwarzschild solution.

This paper is organized as follows.
In Sec. II, we briefly review $f(Q)$ theory, where the action and the equations of motion are introduced.
In Sec. III, we derive the nonmetricity scalar $Q$ and the equations of motion for generic static and spherically symmetric geometry with an anisotropic fluid.
We see that the off-diagonal component of field equations leads to two solutions: STGR which recovers GR, and the constant nonmetricity scalar.
Then, with a focus on the second solution in Sec. IV, we investigate the constraints on the functional form of $f(Q)$ and geometry under various conditions.
Finally, Sec. V is devoted to the conclusions and discussion.

\section{$f(Q)$ gravity} \label{Sec: f(Q) gravity}

$f(Q)$ gravity considers a generic metric-affine spacetime,
in which the metric tensor $g_{\mu\nu}$ and connection $\Gamma_{\mu\nu}^\lambda$ are treated independently, and the nonmetricity of the connection is defined by
\begin{align}
    Q_{\alpha \mu \nu }
    \equiv \nabla _{\alpha }g_{\mu \nu}
    =\partial_{\alpha}g_{\mu \nu} 
    - \Gamma^{\lambda}_{\ \alpha\mu}g_{\lambda \nu} - \Gamma^{\lambda}_{\ \alpha\nu}g_{\mu \lambda}
    \, .
\end{align}
The general form of affine connection can be decomposed into the following three independent components:
\begin{align}
\label{Eq: connect-decomp}
\Gamma^{\lambda}_{\ \mu\nu} =
\left \{ {}^{\lambda}_{\ \mu\nu} \right \} +
K^{\lambda}_{\ \mu\nu}
+L^{\lambda}_{\ \mu\nu} 
\, ,
\end{align}
where $\left\{ {}^{\lambda}_{\ \mu\nu} \right\}$ denotes the Levi-Civita connection determined by the metric $g_{\mu\nu}$,
\begin{align}
\left \{ {}^{\lambda}_{\ \mu\nu} \right \} 
\equiv \frac{1}{2} g^{\lambda \beta} 
\left( \partial_{\mu} g_{\beta\nu} + \partial_{\nu} g_{\beta\mu} - \partial_{\beta} g_{\mu\nu} \right) 
\, .
\end{align}
$K^{\lambda}_{\ \mu\nu}$ is the contortion written as
\begin{align}
K^{\lambda}_{\ \mu\nu} \equiv 
\frac{1}{2} T^{\lambda}_{\ \mu\nu}+T_{(\mu \ \nu)}^{\ \lambda} 
\, ,
\end{align}
with the torsion tensor $T^{\lambda}_{\ \mu\nu}$ defined as the antisymmetric part of the affine connection,
$T^{\lambda}_{\ \mu\nu}\equiv 2 \Gamma^{\lambda}_{\ [\mu\nu]}$,
and the disformation $L^{\lambda}{}_{\mu\nu}$ is defined by
\begin{align}
L^{\lambda}_{\ \mu\nu} \equiv 
\frac{1}{2} Q^{\lambda}_{\ \mu\nu}-Q_{(\mu \ \nu)}^{\ \lambda} 
\, .
\end{align}
Then let us introduce the following nonmetricity conjugate
\begin{align}\label{Eq: nms}
P^{\alpha }_{\ \mu \nu } &= 
-\frac{1}{4}Q^{\alpha}_{\ \mu \nu } +\frac{1}{2}Q_{(\mu \ \nu )}^{\ \  \alpha} 
+\frac{1}{4}\left( Q^{\alpha } -\tilde{Q}^{\alpha }\right)g_{\mu \nu }
-\frac{1}{4}\delta^{\alpha}_{\ (\mu} Q_{\nu)}
\, ,
\end{align}
and its two independent traces,
\begin{align}
    Q_{\alpha}\equiv Q_{\alpha \ \mu}^{\ \mu}
    \,,\quad
    \tilde{Q}_{\alpha } \equiv Q^{\mu}_{\ \alpha \mu }
    \,.
\end{align}
Finally, the nonmetricity scalar is defined as follows
\begin{align}
Q=-Q_{\alpha \mu \nu }P^{\alpha \mu \nu}
\,.
\label{nms}
\end{align}

Supplemented with Lagrange multipliers, one introduces $f(Q)$ gravity given by the following action~\cite{BeltranJimenez:2018vdo}:
\begin{align} 
    \label{Eq: action}
    S=\int \sqrt{-g}~d^4x
    \left[\frac{1}{2}f(Q) + \lambda_{\alpha}^{\ \beta\mu\nu}R^{\alpha}_{\ \beta\mu\nu} + \lambda_{\alpha}^{\ \mu\nu}T^{\alpha}_{\ \mu\nu} + \mathcal{L}_m\right]
    \, ,
\end{align}
where $g$ is the determinant of the metric $g_{\mu\nu}$,
$f(Q)$ is an arbitrary function of the nonmetricity $Q$,
$\lambda_{\alpha}^{\ \beta\mu\nu}$ is the Lagrange multipliers,
and ${\cal L}_m$ is the matter Lagrangian density,

Varying the action~\eqref{Eq: action} with respect to the metric gives the field equation
\begin{align}
\label{Eq: metric-eom}
- T_{\mu\nu} &=
\frac{2}{\sqrt{-g}}\nabla_{\alpha} \left(\sqrt{-g} f_{Q} P^{\alpha}_{\ \mu\nu} \right)
+ \frac{1}{2}g_{\mu\nu} f  
+ f_Q \left( P_{\mu\alpha\beta}Q_{\nu}^{\ \alpha\beta}
-2Q_{\alpha\beta\mu}P^{\alpha\beta}_{\ \ \ \nu}\right)
\,,
\end{align}
where a subscript $Q$ stands for a derivative of $f(Q)$ with respect to $Q$, 
$f_Q \equiv \partial_{Q}f(Q)$. 
The energy-momentum tensor is defined in the standard way,
\begin{align}
    \label{Eq: emt}
    T_{\mu \nu } \equiv 
    -\frac{2}{\sqrt{-g}} \frac{\delta \left(\sqrt{-g} \mathcal{L}_{m}\right)}{\delta g^{\mu \nu}}
    \, . 
\end{align}

Varying Eq.~\eqref{Eq: action} with respect to the connection, one obtains
\begin{align}\label{con-eom}
\nabla_{\rho}\lambda_{\alpha}^{\ \nu\mu\rho}+\lambda_{\alpha}^{\ \mu\nu}
=\sqrt{-g} f_{Q} P^{\alpha}_{\ \mu\nu} + H_{\alpha}^{\ \mu\nu}
\,,
\end{align}
where the hypermomentum tensor density is written as
\begin{align}
\label{Eq: hypermomentum}
H_{\alpha}^{\ \mu\nu} &= 
-\frac{1}{2} \frac{\delta  \mathcal{L}_{m}}{\delta \Gamma^{\alpha}_{\ \mu \nu}}
\,.
\end{align}
By taking into account the antisymmetry property of $\mu$ and $\nu$ in the Lagrangian multiplier coefficients, Eq.~\eqref{con-eom} can be reduced to 
\begin{align}
\nabla_\mu\nabla_\nu \left(\sqrt{-g} f_Q P^{\mu\nu}_{\ \ \alpha}+H_{\alpha}^{\ \mu\nu} \right) &= 0
\, .
\end{align}
Taking $\nabla_\mu\nabla_\nu H_{\alpha}^{\ \mu\nu}=0$ 
(see discussion in Ref.~\cite{BeltranJimenez:2018vdo}),
we have
\begin{align}
\label{Eq: conenction-eom}
\nabla_\mu\nabla_\nu \left(\sqrt{-g} f_Q P^{\mu\nu}_{\ \ \alpha} \right) = 0\,. 
\end{align}
Without curvature and torsion make the affine connection has the following form ~\cite{BeltranJimenez:2017tkd}
\begin{align}
    \Gamma^{\alpha}_{\ \mu\nu} &=\left( \frac{\partial x^{\alpha}}{\partial \xi^{\lambda}} \right) \partial _{\mu} \partial_{\nu}\xi^{\lambda}
    \label{affine-connection}
    \,.
\end{align}
We can make a special coordinate choice, the so-called coincident gauge, so that $\Gamma^{\alpha}_{\ \mu\nu}=0$. Then, the nonmetricity reduces to
\begin{align}
Q_{\alpha \mu \nu } &=\partial_{\alpha}g_{\mu \nu}
\,,
\end{align}
and thereby largely simplifies the calculation since only the metric is the fundamental variable.

However, the cost is that the action no longer remains diffeomorphism invariant, except for STGR~\cite{BeltranJimenez:2019tme}.
One can utilize the covariant formulation of $f(Q)$ gravity to avoid the problem.
Since the affine connection in Eq.~\eqref{affine-connection} is purely inertial, one could utilize the covariant formulation by first determining the affine connection in the absence of gravity~\cite{Zhao:2021zab}.
As shown in this work, however, the off-diagonal component of the field equations in the coincident gauge would put strict constraints on $f(Q)$ gravity, thereby providing us with nontrivial functional forms of $f(Q)$.

\section{Static and Spherically Symmetric Geometry in A Polar Coordinate} \label{Sec: SSS}
The metric ansatz for a generic static and spherically symmetric spacetime is written as
\begin{align}
    \label{Eq: spherical-ansatz}
    ds^{2}=-e^{a(r)}dt^{2}+e^{b(r)}dr^{2} + r^{2}d\Omega^{2}
    \,,
\end{align}
where $d\Omega^{2}\equiv d\theta^{2}+\sin^{2}\theta d\phi^{2}$.
By substituting Eq.~\eqref{Eq: spherical-ansatz} into Eq.~\eqref{nms},
the nonmetricity scalar $Q$ is written in terms of $r$,
\begin{align}
\label{Eq: spherical-nms}
Q(r) &= -\frac{2e^{-b}}{r}\left(a^{\prime}+\frac{1}{r}\right)
\, ,
\end{align}
where a prime ($^{\prime}$) denotes a derivative with respect to the radial coordinate $r$. 

Corresponding to the spherically symmetric geometry,
the energy-momentum tensor for an anisotropic fluid with spherical symmetry is given by
\begin{align}
    T_{\mu \nu}=\left( \rho +p_t \right) u_{\mu}u_{\nu}+p_tg_{\mu \nu}+\left( p_r-p_t \right) v _{\mu}v _{\nu}
    \label{Eq: fluid}
    \, .
\end{align}
Here, $u_{\mu}$ is the four-velocity, and
$v _{\mu}$ is the unitary spacelike vector in the radial direction satisfying $u^\mu u_\mu =-1$, $v^\mu v_\mu=1$, and $u^\mu v_\nu = 0$.
$\rho(r)$ is the energy density,
$p_r(r)$ is the pressure in the direction of $v _{\mu}$ (radial pressure),
and $p_t(r)$ is the pressure orthogonal to $v _{\mu}$ (tangential pressure). 

For the anisotropic fluid \eqref{Eq: fluid}, independent components of equations of motion \eqref{Eq: metric-eom} are listed as follows:
\begin{align}
\label{Eq: eom-dens}
\rho &= \frac{f}{2} - f_Q\left( Q+\frac{1}{r^2}+\frac{e^{-b}}{r}(a^{\prime}+b^{\prime})\right)
\,, \\
\label{Eq: eom-presr}
p_{r} &=- \frac{f}{2} + f_Q\left(Q+\frac{1}{r^2}\right)
\, , \\
\label{Eq: eom-prest}
p_{t} &= -\frac{f}{2} + f_Q\left\{ \frac{Q}{2}-e^{-b}\left[\frac{a^{\prime\prime}}{2}+\left(\frac{a^{\prime}}{4}+\frac{1}{2r}\right) (a^{\prime}-b^{\prime})\right]\right\}
\, , \\
\label{Eq: eom-impos}
0&=\frac{\cot\theta}{2}Q^{\prime}f_{QQ}
\, .
\end{align}

Substituting the nonmetricity scalar Eq.~\eqref{Eq: spherical-nms} into the equations of motion, we have
\begin{align}
    2p_{r}^{\prime}+a^{\prime}\left(\rho+p_{r}\right)
    =f_{Q}\frac{e^{-b}}{r}\left[\frac{2}{r}\left(a^{\prime}+b^{\prime}\right)-a^{\prime}\left(a^{\prime}-b^{\prime}\right)+\frac{4}{r^{2}}\left(1-e^{b}\right)-2a^{\prime\prime}\right]
    =0
    \,,
\end{align}
where in the last equality we have assumed an isotropic fluid and used $p_r-p_t=0$ in Eqs.~\eqref{Eq: eom-presr} and~\eqref{Eq: eom-prest}.
Therefore, the Tolman-Oppenheimer-Volkoff equation is satisfied in general $f(Q)$ gravity for a static and spherically symmetric metric with isotropic fluids.

Now looking at the off-diagonal component in Eq.~\eqref{Eq: eom-impos},
one finds that the solutions to $f(Q)$ gravity are constrained to the following two cases:
\begin{align}
    \label{lin}
    f_{QQ} &= 0 \Rightarrow f(Q)=a_0+a_1Q
    \,, \\
    \label{Q0}
    Q^{\prime} &= 0 \Rightarrow Q=Q_0
    \,,
\end{align}
where $Q_{0}$, $a_{0}$, and $a_{1}$ are constant.
This result is similar to that in $f(T)$ gravity~\cite{Boehmer:2011gw},
i.e. $f_{TT}=0$ or $T'=0$ for the static and spherically symmetric assumption with the diagonal tetrad in $f(T)$ gravity.

The first solution in Eq.~\eqref{lin} is obviously reduced to STGR and is thereby equivalent to GR,
with the ratio $a_{0}/a_{1}$ corresponding to cosmological constant $\Lambda$. 
Nevertheless, we shall confirm whether the Schwarzschild solution exists in linear $f(Q)$ gravity.

In vacuum with $\rho = p_{r} = p_{t} = 0$, the equations of motion are reduced to
\begin{align}
    \label{lin-vac-rho}
    0 &= a^{\prime}+b^{\prime}
    \, ,\\
    \label{lin-vac-pr}
    Q &= \frac{a_0}{a_1} - \frac{2}{r^2}
    \, ,\\
    \label{lin-vac-pt}
    0 &= \frac{a_0}{2} + a_1 e^{-b}\left[ \frac{a^{\prime\prime}}{2}+\left(\frac{a^{\prime}}{4}+\frac{1}{2r}\right) (a^{\prime}-b^{\prime}) \right]
    \, .
\end{align}
The first equation indicates that $a(r)=-b(r)+c$,
where $c$ is an integration constant and can be ignored by rescaling the time coordinate $t$ to $e^{-c/2}t$.
As a result, the $rr$ component is the inverse of the $tt$ component in Eq.~\eqref{Eq: spherical-ansatz}.
The second equation implies a cosmological constant term $2\Lambda=a_{0}/a_{1}$.
Since $Q = - R$ up to a total derivative or surface term in the action of STGR,
the sign of $\Lambda$ is flipped compared with the case in GR due to our convention of nonmetricity scalar Eq.~\eqref{nms}.

Using Eqs.~\eqref{Eq: spherical-nms},~\eqref{lin-vac-rho}, and~\eqref{lin-vac-pr}, we obtain
\begin{align}\label{lin-vac-b}
    e^{-b} =  1 + \frac{c_{1}}{r} -\frac{a_{0}}{6 a_{1}} r^{2}
    \,.
\end{align}
Note that, throughout the paper, we shall use $c_{i}$ to denote the integration constant. Then, the line element in Eq.~\eqref{Eq: spherical-ansatz} is written as
\begin{align}
    \label{lin-vac-metric}
    ds^{2}=
    -\left( 1 + \frac{c_{1}}{r} - \frac{a_0}{6a_1} r^{2} \right) dt^{2}
    +\left( 1 + \frac{c_{1}}{r} - \frac{a_0}{6a_1} r^{2} \right)^{-1} dr^{2}
    + r^{2}d\Omega^{2}
    \,.
\end{align}
One recognizes that it represents the Schwarzschild (anti-)de Sitter solution,
as that in GR,
with $\Lambda=a_{0}/(2a_{1})$ and $c_{1} = -2M$,
where $M$ is a mass parameter.

In the next section, we shall further discuss the nontrivial solution that $Q= Q_{0}$ in Eq.~\eqref{Q0}, which will also lead to strict constraints on the functional forms.
In addition, the constant nonmetricity scalar $Q_0$ could be interpreted as cosmological constant $\Lambda$, as shown in Eq.~\eqref{lin-vac-pr}.

\section{Constant NonMetricity Scalar: $Q=Q_0$}
\label{Sec: Constant}

For constant $Q$, the geometric quantities are related to each other according to Eq.~\eqref{Eq: spherical-nms}
\begin{align}
    \label{Q0-a}
    a^{\prime}&=-\frac{Q_{0}r}{2}e^{b}-\frac{1}{r}
    \,,
\end{align}
and Eqs.~\eqref{Eq: eom-dens}-\eqref{Eq: eom-prest} become
\begin{align}
    \label{Q0-rho}
    \rho&=\frac{f\left(Q_{0}\right)}{2}-f_{Q}\left(Q_{0}\right)\left[Q_{0}+\frac{1}{r^{2}}+\frac{e^{-b}}{r}\left(a^{\prime}+b^{\prime}\right)\right]
    \,, \\
    \label{Q0-pr}
    p_{r} &=- \frac{f(Q_{0})}{2} + f_Q(Q_{0})\left(Q_{0}+\frac{1}{r^2}\right)
    \, , \\
    \label{Q0-pt}
    p_{t} &= - \frac{f(Q_{0})}{2} + f_Q(Q_{0})\left\{ \frac{Q_{0}}{2}-e^{-b}\left[\frac{a^{\prime\prime}}{2}+\left(\frac{a^{\prime}}{4}+\frac{1}{2r}\right) (a^{\prime}-b^{\prime})\right]\right\}
    \, . 
\end{align}
With these four equations in hand, we will try to determine the remaining six quantities, i.e., $f(Q)$, $a(r)$, $b(r)$, $\rho(r)$, $p_{r}(r)$, and $p_{t}(r)$, by imposing some conditions.

\subsection{Vacuum solution}

We first focus on the vacuum case with $\rho = p_r = p_t = 0$, then the equations reduce to
\begin{align}
    \label{Q0-vac-rho}
    0&=f_{Q}\left(Q_{0}\right)\frac{e^{-b}}{r}\left(a^{\prime}+b^{\prime}\right)
    \,,\\
    \label{Q0-vac-pr}
    0&=-\frac{f\left(Q_{0}\right)}{2}+f_{Q}\left(Q_{0}\right)\left(Q_{0}+\frac{1}{r^{2}}\right)
    \,,\\
    \label{Q0-vac-pt}
    0&=f_{Q}\left(Q_{0}\right)\left\{ \frac{Q_{0}}{2}+\frac{1}{r^{2}}+e^{-b}\left[\frac{a^{\prime\prime}}{2}+\left(\frac{a^{\prime}}{4}+\frac{1}{2r}\right)\left(a^{\prime}-b^{\prime}\right)\right]\right\}
    \,.
\end{align}
One immediately finds that from Eq.~\eqref{Q0-vac-pr}
\begin{align}
    \label{Q0-vac-restriction}
    f_{Q}\left(Q_{0}\right) = 0 \, , \ f(Q_{0}) = 0 \, .
\end{align}
Those two restrictions imply that a general functional form of $f(Q)$ should be 
\begin{align}
\label{Q0-vac-model}
    f(Q)=\underset{n}{\sum}a_{n}\left(Q-Q_{0}\right)^{n} \, ,
\end{align}
where $a_{n}$ are parameters.
Therefore, in order for $f(Q)$ gravity to have nontrivial spacetime solutions, the functional form of $f(Q)$ should satisfy Eq.~\eqref{Q0-vac-restriction}, otherwise, we only find solutions in the general relativity, where $f(Q)$ gravity is reduced to STGR as discussed in the previous section.
We will revisit the above and briefly discuss the model building in Conclusions.
In the following, we analyze the spacetime structure, which is less constrained due to the triviality of Eqs.~\eqref{Q0-vac-rho} and~\eqref{Q0-vac-pt} provided that $f(Q_0)=f_Q(Q_0)=0$.

\subsubsection{Case 1: $a'+b'=0$}

Equation~\eqref{Q0-vac-rho} also allows for a solution that $a'+b'=0$,
which is necessary if the Schwarzschild solution is to exist as in Eq.~\eqref{lin-vac-rho}.
However, Eq.~\eqref{Q0-a} solves $b(r)$ as
\begin{align}
\label{Q0-vac-b}
    e^{-b(r)} = \frac{c_{3}}{r}-\frac{Q_{0}}{6}r^2
    \, ,
\end{align}
and the line element Eq.~\eqref{Eq: spherical-ansatz} is then given by
\begin{align}
    \label{Q0-vac-metric}
    ds^2=-\left( \frac{c_3}{r}-\frac{Q_0}{6}r^2 \right) dt^2 
    +\left( \frac{c_3}{r}-\frac{Q_0}{6}r^2 \right) ^{-1}dr^2
    +r^{2}d\Omega^{2}
    \, .
\end{align}
It shows that there is no exact Schwarzschild solution for nontrivial $f(Q)$ functions.
It is, however, a Schwarzschild (anti-)de Sitter-like solution with $c_{3} = -2M$ and $Q_{0}=2\Lambda$,
and its asymptotic behavior shall be the same as Eq.~\eqref{lin-vac-metric} at $r\ll1$ and $r\gg1$.
Note that inversely, substituting the Schwarzschild metric into Eq.~\eqref{Q0-a}, $Q_{0}$ cannot be constant, which guarantees the absence of the exact Schwarzschild solution.

Interestingly, Eq.~\eqref{Q0-vac-metric} is similar to the Schwarzschild-like solution found by the Noether symmetry approach in $f(T)$ gravity ~\cite{Paliathanasis:2014iva},
where the line element is expressed as follows:
\begin{align}
    ds^2=- A(r) dt^2 + \frac{1}{d^{2}_{3}} \frac{1}{A(r)} dr^{2}
    +r^{2}d\Omega^{2}
    \, 
\end{align}
with 
\begin{align}
    A(r) = \frac{2d_{1}}{3d_{3}}r^{2} - \frac{2d_{\mu}}{d_{3}r}.
\end{align}
Here, $d_{\mu} = d_{1}d_{4} - d_{2}d_{3}$, 
and $d_{1} \cdots d_{4}$ are free parameters (in Ref.~\cite{Paliathanasis:2014iva}, they are integration constants).
Identifying $d_{3}=1$, $d_{1} = -Q_{0}/4$, and $d_{\mu} = - c_{3}/2$,
we can rewrite Eq.~\eqref{Q0-vac-metric} in the following form: 
\begin{align}\label{Eq: metric-sec5-2}
     \frac{c_3}{r}-\frac{Q_0}{6}r^2 =\eta \left(1-\frac{r_{\star}}{r}\right)R\left(r\right)
     \, .
\end{align}
In the above expression we have defined
\begin{align}
    \eta &=\left(-\frac{Q_{0}c_{3}^{2}}{6}\right)^{1/3}
    \, , \\
    r_{\star} &=\left(\frac{6c_{3}}{Q_{0}}\right)^{1/3}
    \, , \\
    R(r) &=1+\frac{r}{r_{\star}}+\left(\frac{r}{r_{\star}}\right)^{2}
    \, .
\end{align}
The $r_{\star}$ is a characteristic radius with the restriction $c_{3}Q_{0}>0$ so that  $r_{\star}>0$, 
and the function $R(r)$ can be viewed as a distortion factor
which quantifies the deviation from the exact Schwarzschild solution.
By definition of $\eta>0$, $c_{3}<0$, and thus, $Q_{0}<0$.
Therefore, at the scale $r \rightarrow r_{\star}$, 
Eq.~\eqref{Q0-vac-metric} can mimic the Schwarzschild solution,
and the $f(Q)$ gravity approximately produces a Schwarzschild spacetime in vacuum.

Next, we shall verify whether the term inside the braces in Eq.~\eqref{Q0-vac-pt} can vanish under the condition that $a'+b'=0$.
Substituting Eq.~\eqref{Q0-a} leads to
\begin{align}
0 &= \left( e^{-b} \right)^{\prime \prime} + \frac{2\left( e^{-b} \right)^{\prime}}{r} 
+ \frac{2}{r^2} + Q_{0}
\, ,
\end{align}
which solves $b$ as
\begin{align}
e^{-b(r)} = \left( c_{4} + \frac{c_{5}}{r} - \frac{Q_{0} r^{2}}{6} - 2\ln r \right)
\, .
\end{align}
It is apparently in conflict with Eq.~\eqref{Q0-vac-b}, which indicates the inconsistency between the two possibilities.

\subsubsection{Case 2: $a'+b'\ne 0$}

We could assume $a'+b'\ne 0$ and make the term inside the braces in Eq.~\eqref{Q0-vac-pt} vanish, which reads
\begin{align}
    \label{Eq: eom-sec5b-1}
    \frac{Q_{0}}{2} + \frac{1}{r^2}  -e^{-b}\left[ \frac{a^{\prime\prime}}{2} + \left(\frac{a^{\prime}}{4}+\frac{1}{2r}\right) (a^{\prime}-b^{\prime})\right] = 0
    \,.
\end{align}
Substituting Eq.~\eqref{Q0-a} into the above equation gives
\begin{align}
    \label{Eq: eom-sec5-4}
    Q_{0}^{2} r^{4} + 2Q_{0}r^{2}(2B + B^{\prime}r) + 4 B (B + 4 + B^{\prime}r) = 0
    \,,
\end{align}
where we have defined $B(r) \equiv e^{-b(r)}$.

Although there is no analytic solution,
we can still qualitatively analyze its asymptotic behavior.
Equation~\eqref{Eq: eom-sec5-4} gives $B^{\prime}$ in terms of $B$ and $r$:
\begin{align}
\label{Eq: eom-sec5-4a}
    B^{\prime} &= 
    - \frac{ \left( Q_{0} r^{2} + 2B\right)^{2} + 16 B}{2r\left(Q_{0}r^{2} + 2B \right)} 
    \, .
\end{align}
Let us study the asymptotic form of $B(r)$ at $r \rightarrow 0$ and $r \rightarrow \infty$
by rewriting Eq.~\eqref{Eq: eom-sec5-4a} as
\begin{align}
\label{Eq: eom-sec5-4b}
    B^{\prime} &= 
    - \frac{Q_{0}r}{2} 
    - \frac{B}{r}
    - \frac{8B}{ Q_{0}r^{3} + 2Br}
    \, .
\end{align}

First, we consider the case $r \rightarrow 0$.
Assuming $B$ converges, 
Eq.~\eqref{Eq: eom-sec5-4b} approximates $B^{\prime} \sim - B/r - 4/r$.
The solution is $B(r) \sim -4 + c_{6}/r$, but it is inconsistent with the assumption.
Thus, $B$ should diverge at $r \rightarrow 0$, 
and then, the second term in Eq.~\eqref{Eq: eom-sec5-4b} is dominant, 
$B^{\prime} \sim - B/r$.
Then, we find the asymptotic form of $b(r)$ at $r \sim 0$:
\begin{align}
\label{Eq: eom-sec5-4b-1}
    e^{-b(r)} \sim \frac{c_{7}}{r}
    \, .
\end{align}

Next, we consider the case $r \rightarrow \infty$.
In a way similar to the case $r\rightarrow 0$,
assuming $B$ converges to a finite value at $r \rightarrow \infty$,
the first term in Eq.~\eqref{Eq: eom-sec5-4b} is dominant,
which gives an inconsistent result $B(r)\sim -Q_{0}r^{2}/4 + c_{8}$.
Thus, $B(r)$ should diverge also at $r \rightarrow \infty$.
Assuming $B \sim r^{n}$, 
the case $0<n<2$ suggests the first term in Eq.~\eqref{Eq: eom-sec5-4b} is dominant, which gives $B(r) \sim -Q_{0}r^{2}/2n$, and it is inconsistent with the assumption.
Moreover, the case $n>2$ suggests the second term is dominant,
which also gives the inconsistent result $B \sim c_{9}/r$.
Finally, one finds a consistent form $B(r) \sim B_{2} r^{2}$,
where $B_{0}$ is a parameter.
In this case, Eq.~\eqref{Eq: eom-sec5-4b} approximates $B^{\prime} \sim - Q_{0}r/2 - B/r$.
Therefore, the coefficient $ B_{2}$ satisfies $2B_{2} = - Q_{0}/2 - B_{2}$, to find $B_{2} = - Q_{2}/6 $.
Finally, we find the asymptotic form of $b(r)$ at $r \sim \infty$:
\begin{align}
\label{Eq: eom-sec5-4b-2}
e^{-b(r)} \sim - \frac{Q_{0}}{6} r^{2}
\, .
\end{align}

The above analysis of the asymptotic behavior suggests that 
the solution $e^{-b(r)}$ to Eqs.~\eqref{Eq: eom-sec5b-1} and~\eqref{Q0-a} 
is similar to the Schwarzschild-like solution in Eq.~\eqref{Q0-vac-b}.
However, $a^{\prime} + b^{\prime} \neq 0$,
and $a(r)$ is determined by Eq.~\eqref{Q0-a}.
$e^{-b(r)} \sim c_{10}/r$ at $r\sim0$,
and then, Eq.~\eqref{Q0-a} gives $a^{\prime}\sim - Q_{0} r^{2}/2c_{10}  - 1/r$.
Because the second term is dominant,
the asymptotic form of $a(r)$ at $r\sim0$ is
$e^{a(r)} \sim c_{11}/r$.
In the same manner, $e^{-b(r)} \sim - \frac{Q_{0}}{6} r^{2}$ at $r\sim \infty$,
and Eq.~\eqref{Q0-a} approximates $a^{\prime}\sim 2/r$.
Thus, the asymptotic form of $a(r)$ at $r\sim \infty$ is
$e^{a(r)} \sim c_{12}r^{2}$.

\subsubsection{Case 3: $e^{a}=1-\frac{2M}{r}$}

In previous case studies, we have confirmed the Schwarzschild-like solution and its asymptotic behavior.
Presupposing that $e^{a}=1-\frac{2M}{r}$, 
we further discuss the deviation from the exact Schwarzschild metric,
where the deviation is written only by the $rr$ component of metric $e^{b(r)}$.
Under the above assumption, Eq.~\eqref{Q0-a} solves $b$ as
\begin{align}
    e^{b}=-\frac{2}{Q_{0}r\left(r-2M\right)}
    \,,
\end{align}
where we require $Q_{0}<0$, and the metric reads
\begin{align}\label{Q0-vac-cas-metric}
    ds^{2}=-\left(1-\frac{2M}{r}\right)dt^{2}+\frac{-2}{Q_{0}r^{2}}\left(1-\frac{2M}{r}\right)^{-1}dr^{2}+r^{2}d\Omega^{2}
    \,.
\end{align}
Comparing the above with the Schwarzschild metric, one finds an extra prefactor in the $rr$ component. 
Taking the limit $r\gg 2M$, 
we find the leading term in the $rr$ component is proportional to $1/r^{2}$,
which shows the absence of a conventional Newtonian limit.

\subsection{Nonvacuum solutions}

We have seen that the vacuum solution for $Q=Q_0$ is closely related to the solutions of Eq.~\eqref{Q0-vac-restriction}, 
which provide us with the condition for the nontrivial functional forms of $f(Q)$.
In this subsection, however, we shall focus on the nonvacuum case by imposing conditions on geometric quantities. 
In particular, we assume that $f_{Q}(Q_0)\ne 0$ to look for particular solutions of interest.

\subsubsection{Case 1: $a'+b'=0$}

The solutions for $a'+b'=0$ are the same as Eqs.~\eqref{Q0-vac-b} and~\eqref{Q0-vac-metric}, while the equations of motion are constrained to
\begin{align}
    -\rho& = p_{r} = p_{t}+f_{Q}\left(Q_{0}\right)\frac{1}{r^{2}}
    \,,\\
    p_{t}&=-\frac{f\left(Q_{0}\right)}{2}+Q_{0}f_{Q}\left(Q_{0}\right)
    \,.
\end{align}
Note that the energy density and radial pressure obey the inverse-square law with respect to the radius if $f_{Q} (Q_{0}) \neq0$, although the tangential pressure is constant. And since $p_{t}$ is a constant in terms of $Q_0$, we rewrite it as $p_{t0}$ and obtain a particular solution to $f(Q)$:
\begin{align}
    f(Q)=c_{13}Q_{0}\sqrt{\frac{Q}{Q_{0}}}-2p_{t0}
    \,.
\end{align}

\subsubsection{Case 2: $a'=0$}
When $a$ is a constant, Eq.~\eqref{Q0-a} solves $b(r)$ as
\begin{align}
    e^{-b}=-\frac{Q_{0}}{2}r^{2}
    \,.
\end{align}
Then the metric reads
\begin{align}
    ds^{2}=-dt^{2}-\frac{2}{Q_{0}r^{2}}dr^{2} + r^{2}d\Omega^{2}
    \,,
\end{align}
where we have absorbed $a$ into a time coordinate. 
The equations of motion are
\begin{align}
    \rho&=-p_{r}-Q_{0}f_{Q}\left(Q_{0}\right)
    \,,\\
    p_{r}&=p_{t}+f_{Q}\left(Q_{0}\right)\frac{1}{r^{2}}
    \,,\\
    p_{t}&=-\frac{f\left(Q_{0}\right)}{2}+Q_{0}f_{Q}\left(Q_{0}\right)
    \,.
\end{align}
The values of the pressures are the same as in case 1, so is the particular solution to $f(Q)$.

\subsubsection{Case 3: $b=0$}

For $b=0$, Eq.~\eqref{Q0-a} gives
\begin{align}
    e^{a}=\frac{c_{14}}{r}\exp\left(-\frac{Q_{0}}{4}r^{2}\right)
    \,,
\end{align}
with the metric being
\begin{align}
    ds^{2}=-\frac{c_{14}}{r}\exp\left(-\frac{Q_{0}}{4}r^{2}\right)dt^{2}+dr^{2}
    +r^{2}d\Omega^{2}
    \,.
\end{align}
Then the equations of motion read
\begin{align}
    \rho&=\frac{f\left(Q_{0}\right)}{2}-\frac{Q_{0}f_{Q}\left(Q_{0}\right)}{2}
    \,,\\
    p_{r}&=-\rho+f_{Q}\left(Q_{0}\right)\left(\frac{Q_{0}}{2}+\frac{1}{r^{2}}\right)
    \,,\\
    p_{t}&=-\rho-f_{Q}\left(Q_{0}\right)\left(\frac{Q_{0}}{4}+\frac{Q_{0}^{2}r^{2}}{16}+\frac{1}{4r^{2}}\right)
    \,.
\end{align}
Since $\rho$ is a constant in terms of $Q_0$, we rewrite it as $\rho_0$. Then, a particular solution to $f(Q)$ is
\begin{align}
    f(Q)=c_{15} Q + 2\rho_{0}
    \,,
\end{align}
which recovers STGR with cosmological constant $\Lambda=\rho_{0}/c_{15}$.

\section{Conclusions and Discussion} \label{Sec: Conclusion}

Different geometric bases allow us to have different descriptions of gravity.
Of particular concern in recent years is the STGR, in which gravity is attributed to the nonmetricity tensor. 
As an extension of STGR, $f(Q)$ gravity is an intriguing approach in the study of modified gravity. 
In this work, we have analyzed the static and spherically symmetric solutions in general $f(Q)$ gravity
and have shown the two possible solutions due to the nonzero off-diagonal components,
that is, the linear $f(Q)$ function and constant nonmetricity scalar $Q=Q_{0}$.
We have confirmed that the Schwarzschild (anti-)de Sitter solution only exists in linear $f(Q)$ gravity,
which recovers STGR and is thereby equivalent to GR.

With a focus on the second solution $Q=Q_0$,
we have found that the vacuum solutions $f(Q_0)=f_Q(Q_0)=0$ as in Eq.~\eqref{Q0-vac-restriction}, 
which puts stringent constraints on the functional form of $f(Q)$ gravity.
On the other hand, those solutions do not constrain the metric components and thereby provide us with considerable freedom to investigate the spacetime structure. 
As examples, we have discussed three cases trying to find the exact Schwarzschild solution,
only to obtain the Schwarzschild-like solutions.
Those solutions do not exhibit asymptotic flatness; that is, 
the spacetime cannot reduce to the Minkowski one at $r\gg2M$ due to the absence of $1$ in the metric. 
Therefore, the Schwarzschild-like solutions obtained for the case $Q=Q_{0}$ cannot recover the Newtonian inverse-square law, which is strictly constrained by observations.
As for the nonvacuum solution for $Q=Q_0$, we have started from the geometric assumptions listed as three examples.
By proper conditions on the $tt$ component or $rr$ component,
we have found some particular solutions to $f(Q)$ with constant energy density or constant tangent pressure.

As an application of our findings, we have discussed the implications for the $f(Q)$ cosmology from the theoretical viewpoint of model building. 
In the case study for $Q=Q_0$,
we have seen that $Q_{0}$ plays a role of the cosmological constant in the Schwarzschild-like solutions,
and thus, we can expect the cosmological model of $f(Q)$ gravity for DE.
The possible functional form of $f(Q)$ is constrained by Eq.~\eqref{Q0-vac-restriction}, which suggests $f(Q)=\underset{n}{\sum}a_{n}\left(Q-Q_{0}\right)^{n}$ as in Eq.~\eqref{Q0-vac-model}.
Therewith, specifying the coefficients $a_{n}$, one could construct specific models of $f(Q)$ gravity.
The simplest example is the polynomial of $Q$, which was actually proposed and investigated as the cosmological model of $f(Q)$ gravity~\cite{BeltranJimenez:2019tme}.
Moreover, reading Eq.~\eqref{Q0-vac-model} as the Taylor expansion, 
one can construct the exponential of $Q$ or its combinations, for instance, the trigonometric functions.
Those functional forms are known well in cosmological applications of $f(R)$ gravity~\cite{DeFelice:2010aj,Starobinsky:2007hu,Appleby:2007vb}, 
and thus, the $f(Q)$ functions similar to or the same as the known $f(R)$ functions can provide us with intriguing cosmology.

In the end, we comment on the gauge fixing and possible analogy to the known results in $f(T)$ gravity.
The curvatureless and torsionless conditions correspond to the coincident gauge so that the affine connection is always zero.
However, the choice of the coincident gauge makes the $f(Q)$ theory no longer invariant under the general coordinate transformation.
As in the case of $f(T)$ gravity,
the covariant formulation of $f(Q)$ theory~\cite{Zhao:2021zab,Lin:2021uqa,DAmbrosio:2021zpm} can allow us to avoid issues caused by the  coincident gauge.
In our present analysis, the existence of the constraint equation~\eqref{Eq: eom-impos} results in two choices for $f(Q)$ gravity: 
a linear function or constant nonmetricity scalar.
Concerning the gauge fixing and breakdown of diffeomorphism,
the different coordinate systems may allow $f(Q)$ gravity to possess the Schwarzschild solution even in the noncovariant approach to $f(Q)$ gravity.
As in the $f(T)$ theory, an off-diagonal component of the field equations shows up in the polar coordinate,
which prohibits the Schwarzschild spacetime as a solution, 
otherwise, the $f(T)$ function should be linear to the torsion scalar $T$.
On the other hand, $f(T)$ gravity with higher orders of $T$ can possess the Schwarzschild solution in the isotropic coordinate~\cite{Ferraro:2011ks}.
By relying on the analogy between $f(T)$ and $f(Q)$ theories, 
it may be worth studying the spherically symmetric solution of $f(Q)$ gravity in the isotropic coordinate.
Applying our present analysis to the covariant $f(Q)$ gravity or other coordinate systems,
we may be able to obtain the Schwarzschild solution even in the nontrivial $f(Q)$ function.

\begin{acknowledgments}
The authors thank Taotao Qiu for valuable comments on the scientific content.
T.K. is grateful to Shin'ichi Nojiri for his advice on the calculations.
\end{acknowledgments}

\bibliographystyle{apsrev4-1}
\bibliography{References}

\end{document}